\documentclass[aps,prl,twocolumn,showpacs,superscriptaddress,floatfix]{revtex4-1}  % for review and submission
\usepackage{graphicx}  % needed for figures
\usepackage{dcolumn}   % needed for some tables
\usepackage{bm}        % for math
\usepackage{amsmath}   % for split env
\usepackage{amssymb}   % for math
\usepackage{color}
\usepackage{overpic,subfigure}

\usepackage[pscoord]{eso-pic}% The zero point of the coordinate system is the lower left corner of the page (the default).

\newcommand{\newsection}[1]{{\itshape #1.---}}

\newcommand{\B}{\boldsymbol{B}}
\newcommand{\C}{{\cal C}}
\renewcommand{\d}{\mathrm{d}}
\newcommand{\dTpara}{\delta T_{\parallel}}
\newcommand{\dTparahat}{\delta \hat{T}_{\parallel}}

\newcommand{\E}{\boldsymbol{E}}
\newcommand{\eg}{\emph{e.g.}}
\newcommand{\fig}{Fig.}
\newcommand{\figs}{Figs.}
\newcommand{\g}{\hat{g}_{m}}
\newcommand{\gp}{\hat{g}_{m+1}}
\newcommand{\gm}{\hat{g}_{m-1}}

\newcommand{\kpara}{k_{\parallel}}
\newcommand{\ppara}{p_{\parallel}}
\newcommand{\qpara}{q_{\parallel}}
\newcommand{\kperp}{k_{\perp}}
\newcommand{\kperpmax}{k_{\perp\max}}
\newcommand{\kperpzero}{k_{\perp0}}

\newcommand{\Lpara}{L_{\parallel}}
\newcommand{\Lperp}{L_{\perp}}

\newcommand{\M}{{\cal M}}
\newcommand{\N}{{\cal N}}
\newcommand{\pd}[2]{\frac{\partial #1}{\partial #2}}
\newcommand{\fd}[2]{\frac{\d #1}{\d #2}}

\renewcommand{\r}{\boldsymbol{r}}

\newcommand{\rhs}{right-hand side}
\newcommand{\rperp}{\r_{\perp}}
\renewcommand{\S}{{\cal S}}
\newcommand{\sgn}{{\mathrm{sgn}}}

\newcommand{\T}{{\cal T}}

\newcommand{\tausi}{\tau_{\mathrm{s}}^{-1}}

\newcommand{\taunli}{\tau_{\mathrm{nl}}^{-1}}
\newcommand{\tfd}[2]{\mathrm{d}#1/{\mathrm{d}#2}}
\newcommand{\tpd}[2]{\partial#1/{\partial#2}}

\renewcommand{\u}{\boldsymbol{u}}
\newcommand{\uperp}{\boldsymbol{u}_{\perp}}
\newcommand{\uperphat}{\hat{\boldsymbol{u}}_{\perp}}
\newcommand{\upara}{u_{\parallel}}
\newcommand{\uparahat}{\hat{u}_{\parallel}}
\newcommand{\vpara}{v_{\parallel}}
\newcommand{\vparahat}{\hat{v}_{\parallel}}

\newcommand{\vperpv}{\boldsymbol{v}_{\perp}}
\newcommand{\vth}{v_{\mathrm{th}}}

\newcommand{\z}{\boldsymbol{z}}

\newcommand{\Wfluid}{W_{\textrm{fluid}}}
\newcommand{\Wkinetic}{W_{\textrm{kin}}}
\newcommand{\ta}[1]{\left\langle#1\right\rangle}
\newcommand{\indthree}{{\cal I}_{m\geq3}}

\renewcommand{\Re}{\mathrm{Re}}
\renewcommand{\Im}{\mathrm{Im}}

\begin{document}

\title{Suppression of phase mixing in drift-kinetic plasma turbulence}
\newcommand{\mathsinst}{\affiliation{OCIAM, Mathematical Institute, University of Oxford, Oxford OX2 6GG, United Kingdom}}
\renewcommand{\mathsinst}{\affiliation{OCIAM, Mathematical Institute, University of Oxford, Andrew Wiles Building,
Radcliffe Observatory Quarter, Woodstock Road, Oxford OX2 6GG, UK}}
  \newcommand{\tp}{\affiliation{Rudolf Peierls Centre for Theoretical Physics, University of Oxford, Oxford OX1 3NP, United Kingdom}}
	\renewcommand{\tp}{\affiliation{Rudolf Peierls Centre for Theoretical Physics, University of Oxford, 1 Keble Road, Oxford OX1 3NP, UK}}
\newcommand{\magdalen}{\affiliation{Magdalen College, University of Oxford}}
\newcommand{\maryland}{\affiliation{Maryland}}
\newcommand{\merton}{\affiliation{Merton College, Merton Street, Oxford OX1 4JD, UK}}
\newcommand{\brasenose}{\affiliation{Brasenose College, Radcliffe Square, Oxford OX1 4AJ, UK}}
\newcommand{\stfc}{\affiliation{Science and Technology Facilities Council, Rutherford Appleton Laboratory, Harwell Campus, Didcot OX11 0QX, UK}}

\author{J. T. Parker} 
\email{joseph.parker@stfc.ac.uk}
\stfc
\mathsinst\brasenose
\author{E. G. Highcock} \brasenose\tp
\author{A. A. Schekochihin} \tp \merton
\author{P. J. Dellar} \mathsinst
\date{\today}

\begin{abstract}
Transfer of free energy from large to small velocity-space scales by phase mixing leads to Landau damping in a linear plasma. In a turbulent drift-kinetic plasma, this transfer is statistically nearly canceled by an inverse transfer from small to large velocity-space scales due to ``anti-phase-mixing'' modes excited by a stochastic form of plasma echo. Fluid moments (density, velocity, temperature) are thus approximately energetically isolated from the higher moments of the distribution function, so phase mixing is ineffective as a dissipation mechanism when the plasma collisionality is small.
%%%Transfer and dissipation of free energy in drift-kinetic plasma turbulence are studied. Transfer of free energy from large to small parallel-velocity-space scales by phase mixing (which in a linear plasma leads to Landau damping) is statistically nearly cancelled by an inverse transfer from small to large velocity scales by ``anti-phase-mixing'' modes excited by the stochastic version of the plasma echo. Consequently, fluid moments (density, fluid velocity, temperature) are approximately energetically isolated from the higher kinetic moments of the distribution function. Very little free energy is transferred to fine velocity-space scales, so parallel phase mixing is ineffective as a dissipation mechanism when the collisionality of the plasma is sufficiently small.
\end{abstract}

\maketitle

\newsection{Introduction}%
Kinetic turbulence in weakly collisional, strongly magnetized plasmas is ubiquitous in magnetic-confinement-fusion experiments \cite{Conner94,Doyle07,Garbet10} and in astrophysical settings \cite{Howes06,Schekochihin09}. Like fluid turbulence, kinetic turbulence may be described as the injection (\eg, by a plasma instability), cascade to small scales, and dissipation of a quadratic invariant, viz., free energy. On spatial scales larger than the ion Larmor radius, kinetic turbulence incorporates two mechanisms for dissipating free energy into heat.
The first is a fluid-like nonlinear cascade from large 
to smaller, sub-Larmor, spatial scales 
(where the free energy is dissipated 
eventually
by collisions \cite{Schekochihin09,Tatsuno09,Howes11,Told15}).
The second is
parallel phase mixing, a linear process 
that transfers free energy from the fluid moments (density, 
fluid velocity
and temperature) 
to the kinetic
(higher-order)
 moments
by creating perturbations in the velocity distribution
on ever finer scales in velocity space, perturbations which are also then dissipated by collisions.
In a linear plasma, this is known as Landau damping and the free energy is dissipated at a rate independent of collision frequency \cite{Zocco11,Kanekar14}.

The macroscopic properties of the turbulence (such as heat and momentum transport) are directly affected by the %interaction of these 
two energy dissipation channels;
yet, while each is understood in isolation, how they interact is not clear.  
Recent work leads to some disquieting observations.
Firstly, a fluid-like theory for the nonlinear cascade \cite{BarnesEtal11} 
predicts power-law spectra for the electrostatic potential in good agreement with those found in gyrokinetic simulations, 
but its derivation neglects free-energy transfer by phase mixing \cite{Schekochihin15}, contrary to what might be expected on the basis of linear theory. Including a constant flux of free energy into velocity space leads to non-universal spectra that tend to be steeper than those empirically observed \cite{Schekochihin15,Howes08,Podesta10,Bratanov13,Passot15,Sulem16}.
Some simulations show a significant proportion of injected free energy cascading and dissipating 
in velocity space \cite{Watanabe06},
albeit with a slower transfer rate than in the linear case,
and with a dissipation rate that depends on collision frequency \cite{HatchEtAl13,Hatch14}.
These observations suggest a complicated relationship between parallel phase mixing and the nonlinear cascade;
there is as yet no complete picture of free-energy flow and dissipation in phase space.

In this Letter, we propose the outlines of such a picture for electrostatic drift-kinetic turbulence. We show that the net transfer of free energy from fluid to kinetic modes is strongly inhibited in a turbulent plasma, compared to a ``linear plasma''. This is due to a stochastic version of the classic plasma-echo phenomenon \cite{Gould67,Malmberg68}: the nonlinearity excites ``anti-phase-mixing'' modes that transfer free energy from small to large velocity-space scales, leading to statistical cancellation of the free-energy flux. The significance of this effect depends on the relative rates of phase mixing and nonlinear advection. We identify regions of wavenumber space where either the echo effect dominates, or phase mixing occurs at the usual linear rate. Most of the free energy contained in fluid moments is at wavenumbers that lie within the echo-dominated region. Therefore, there is very little net free-energy transfer to fine velocity-space scales via linear phase mixing. Consequently, Landau damping is strongly suppressed as a dissipation mechanism.

\newsection{Drift kinetics}%
  We study electrostatic ion-temperature-gradient (ITG) driven drift-kinetic turbulence in an unsheared slab with kinetic ions and Boltzmann electrons.
The equations are the drift-kinetic equation for ions,
\begin{equation}
  \begin{split}
		%\fd{g}{t}
		 \pd{g}{t} 
		+ \vpara\nabla_{\parallel} \left( g + \varphi F_0 \right)
		+ \uperp\cdot\nabla_\perp g
		= C[g] + \chi,
    \label{eq:GyrokineticEquation}
  \end{split}
\end{equation}
and the quasineutrality condition \footnote{We ignore here the subtleties of the $\kpara=0$ electron response in ITG turbulence (see \eg\ \cite[][sec.\ J.2]{Abel13})---they do not matter for the inertial-range physics on which we focus here. The response we use is formally correct for electron-temperature-gradient (ETG) turbulence \cite{Dorland00}, which is described by the same equations with $e\leftrightarrow i$ swapped and some irrelevant sign changes \cite[][sec.\ 2.2.1.]{Schekochihin15}.} 
\begin{equation}
  \begin{split}
		\varphi \equiv \frac{Ze\phi}{T_i} = \alpha \int_{-\infty}^{\infty}\d\vpara ~  g , 
		~ ~ ~ ~ ~ 
		\alpha = \frac{ZT_e}{T_i} .
    \label{eq:Quasineutrality}
  \end{split}
\end{equation}
Here $g=(1/n_i)\int\d^2\vperpv~\delta f$ is the perturbed ion distribution function integrated over perpendicular velocity space, with $n_i$ the mean ion density; $\phi$ is the electrostatic potential, $-e$ the electron charge, $Ze$ is the ion charge, and $T_i$ and $T_e$ the mean ion and electron temperatures;
$F_0(\vpara/\vth)=e^{-\vpara^2/\vth^2}/\sqrt{\pi}$ is the one-dimensional Maxwellian, with $\vpara$ the parallel velocity, $\vth=\sqrt{2T_i/m_i}$ the ion thermal velocity, and $m_i$ the ion mass;
$\uperp=(\rho_i\vth/2)\hat{\z}\times\nabla_{\perp}\varphi$ is the $\E\times\B$ velocity with ion gyroradius $\rho_i$, and $\hat{\z}$ the unit vector in the direction of the magnetic field line.
The perpendicular directions are $x$ and $y$.
The energy-injection term due to a mean ITG in the negative $x$ direction is 
\begin{equation}
  \begin{split}
		\chi=-\frac{\rho_i\vth}{2L_T}\pd{\varphi}{y}\left( \frac{\vpara^2}{\vth^2}-\frac{1}{2}\right) F_0,
\hspace{0.6cm}
\frac{1}{L_T} = - \fd{\ln T_i}{x}.
    \label{eq:Chi}
  \end{split}
\end{equation}
The collision operator $C[g]$ will be described shortly.

The system (\ref{eq:GyrokineticEquation}--\ref{eq:Quasineutrality}) conserves the free energy $W = \int\d^3\r\ \varphi^2/2\alpha + \int\d^3\r \int_{-\infty}^{\infty}\d\vpara ~ g^2/2F_0$,
except for injection by the ITG and dissipation by collisions:
\begin{equation}
  \begin{split}
		\fd{W}{t} = \int\d^3\r\int_{-\infty}^{\infty}\! \d \vpara ~ \frac{g\chi}{F_0} + \int\d^3\r\int_{-\infty}^{\infty}\!\d\vpara~ \frac{gC[g]}{F_0}.
    \label{eq:TotalFreeEnergyConservation}
  \end{split}
\end{equation}

\newsection{Simulations}%
  We study the saturated state of drift-kinetic turbulence 
	in a box with parallel and perpendicular lengths $\Lpara$ and $\Lperp$.
	We solve equations 
	(\ref{eq:GyrokineticEquation}--\ref{eq:Quasineutrality})
	with {\sc SpectroGK} \cite{ParkerThesis,Parker15}, % \cite{SGK}, %
a phase-space-spectral code designed to capture discrete free-energy conservation exactly.
We use a Fourier representation in physical space, $\r$, with 128$\times$128$\times$256 wavenumbers $(k_x,k_y,\kpara)$.
The highest Fourier modes are damped using 8th-order hyperviscosity.
  In the parallel-velocity space, we use the Hermite representation 
	$g(\vpara)=\sum_{m=0}^{\infty}g_mH_m(\vparahat)F_0(\vparahat)/\sqrt{2^mm!}$,
where
	$H_m(\vparahat)=e^{\vparahat^2}(-\mathrm{d}/\mathrm{d}\vparahat)^me^{-\vparahat^2}$
	with ${\vparahat} = \vpara/\vth$.
	For large $m$, %Hermite polynomials behave as %like
	$H_m\sim \cos(\vparahat\sqrt{2m}-m\pi/2)\sqrt{2^mm!}/F_0^{1/2}(\vparahat)$,
	so %each $m$ 
	$\sqrt{m}$ represents a ``wavenumber'' in velocity space.
	The first three Hermite moments are ``fluid'' quantities:
$g_0=\varphi/\alpha$ (density), 
$g_1=\sqrt{2}\upara/\vth$ (parallel fluid velocity),
$g_2=\dTpara/T_i\sqrt{2}$ (parallel-temperature perturbation).
For $m\geq3$, $g_m$ are ``kinetic'' moments,
representing finer velocity-space scales.
We use 256 Hermite modes,
regularizing with 6th-order hypercollisions $C[g_m]=-\nu m^ng_m\indthree$,
where $n=6$, and
$\indthree=1$ if $m\geq3$ and $\indthree=0$ otherwise.
For $n=1$, this is a momentum- and energy-conserving version of the Lenard--Bernstein operator \cite{Kirkwood46,Lenard58}.
As the linear growth rate of the ITG instability in drift kinetics increases indefinitely with $\kperp$, we 
artificially suppress the temperature gradient by a factor 
%$\sim \exp[-200((\kperp/k_{\perp\max})^2+(\kpara/k_{\parallel\max})^2)]$
%$\varkappa = \exp[-200((\kperp/k_{\perp\max})^2+(\kpara/k_{\parallel\max})^2)]$
$\varkappa = \exp[-200(\kperp^2/k^2_{\perp\max}+\kpara^2/k^2_{\parallel\max})]$
to separate the free-energy injection and dissipation scales.
This captures the essential feature of ITG turbulence: that the nonlinear turnover rate eventually dominates the injection rate as $\kperp$ increases.
However, as the 
nonlinear and linear
characteristic rates are 
respectively $\taunli\sim\kperp^{4/3}$ and $\omega^*\sim k_y$, \cite{BarnesEtal11,Schekochihin15}, 
it is necessary to 
limit 
the free-energy injection
artificially to a narrow range of small wavenumbers
to allow an inertial range to develop with the available resolution.
Now only low wavenumbers can grow, 
and only for very large temperature gradients.
We present results for $L_{\parallel}/L_T=1600$, 
with the fastest growing wavenumber $\kperpzero=4\pi/L_{\perp}\approx \kperpmax/30$ setting the energy-injection scale. 
While this arrangement does not inject free energy in a realistic fashion, we are able to study the key features of its transfer and dissipation in drift-kinetic turbulence.

While we solve fully spectrally, it is convenient for presentation to write equations that are spectral in velocity and the parallel spatial direction only.
Equation \eqref{eq:GyrokineticEquation} becomes
\begin{equation}
  \begin{split}
  \pd{\g}{t} 
	&+ i\kpara\vth \left(\sqrt{\frac{m+1}{2}}\gp + \sqrt{\frac{m}{2}} \gm \right)
  \\
	&+ \frac{i\kpara\vth}{\sqrt{2}}\hat{\varphi} \delta_{m1}
	+ \sum_{\ppara+\qpara=\kpara} \hat{\u}_{\perp}(\ppara)\cdot\nabla_{\perp}\g(\qpara)
	\\
    &= 
    -\nu m^n \g\indthree  
    %C[\g]   
    + \hat{\chi},
    \label{eq:GyrokineticEquationFourierHermite}
  \end{split}
\end{equation}
where hats denote functions in $(\rperp,\kpara,m)$ space,
and
$\hat{\chi}=-{\hat{\varkappa}}(\rho_i\vth/2\sqrt{2}L_T)(\tpd{\hat{\varphi}}{y})\delta_{m2}$.
Equation \eqref{eq:GyrokineticEquationFourierHermite} exhibits clearly the two mixing mechanisms present in the turbulence: the ``fluid'' cascade due to the $\uperphat\cdot\nabla_{\perp}\hat{g}_m$ nonlinearity,
and linear phase mixing by coupling between Hermite modes.

\newsection{Free-energy transfer}%
We first formulate the free-energy transfer between fluid and kinetic moments.
Let $W=\int\d^2\rperp\sum_{k\parallel}(\Wfluid+\Wkinetic)$,
where $\Wfluid(\rperp,\kpara)= (1+\alpha)|\hat{\varphi}|^2/2\alpha^2 + |\uparahat|^2/\vth^2 + |\dTparahat|^2/4T^2$ is the free energy in fluid moments,
and $\Wkinetic(\rperp,\kpara)=\sum_{m=3}^{\infty}|\g|^2/2$ is the free energy in kinetic moments.
Equation \eqref{eq:GyrokineticEquationFourierHermite} implies
\begin{equation}
  \begin{split}
	%\fd{}{t} \frac{|\g|^2}{2}
	  \pd{}{t} \frac{|\g|^2}{2}
&	+  \Gamma_m - \Gamma_{m-1} 
	+	\Im\left( \frac{\kpara\vth}{\sqrt{2}}\g\hat{\varphi}^*\delta_{m1}\right)
\\ & 
	+ N_m
    = 
    -\nu m^n |\g|^2\indthree  + \Re \left(\g^*\hat{\chi}\right) ,
    \label{eq:FreeEnergyModeByMode}
  \end{split}
\end{equation}
where 
$N_m=\Re\left[\sum_{\ppara+\qpara=\kpara}\hat{g}_m(\kpara)^*\hat{\u}_{\perp}(\ppara)\cdot\nabla_\perp\hat{g}_m(\qpara)\right]$ is the free-energy transfer in $(\rperp,\kpara)$ space due to the nonlinearity,
and $\Gamma_m=\kpara\vth\sqrt{(m+1)/2}\ \Im(\gp^*\g)$ is the free-energy transfer from mode $m$ to $m+1$ due to phase mixing \cite{Watanabe04}.

Summing \eqref{eq:FreeEnergyModeByMode} separately over $m\in\{0,1,2\}$ (fluid) and $m\geq3$ (kinetic) gives
\begin{equation}
  \begin{split}
%%%		\fd{\Wfluid}{t} = \S - \T, \hspace{1cm}
%%%		\fd{\Wkinetic}{t} = \T - \C,
		\pd{\Wfluid}{t} = \M + \S - \T, \hspace{0.9cm}
		\pd{\Wkinetic}{t} = \N + \T - \C,
    \label{eq:FreeEnergyConservation}
  \end{split}
\end{equation}
where 
$\S(\rperp,\kpara)=\Re(\dTparahat^* \hat{u}_x)/(2T_iL_T)$ 
is the source due to the ITG,
$\C(\rperp,\kpara)=\nu\sum_{m=3}^{\infty}m^n|\g|^2$ is the sink due to collisions,
$\M(\rperp,\kpara)=-\sum_{m=0}^{2}(1+\alpha\delta_{m0})N_m$
and
$\N(\rperp,\kpara)=-\sum_{m=3}^{\infty}N_m$
are free-energy transfers due to the nonlinearity,
and  
%$\T(\rperp,\kpara)=\kpara\vth\sqrt{3}\ \Im\left(\hat{g}^*_3\dTparahat\right)/(2T_i) =\Gamma_2$
$\T(\rperp,\kpara)=\kpara\vth\sqrt{3}\ \Im(\hat{g}^*_3\dTparahat)/(2T_i) =\Gamma_2$
is the transfer of free energy from fluid to kinetic moments due to streaming $\vpara\nabla_{\parallel}g$ in \eqref{eq:GyrokineticEquation}.
Streaming is linear and reversible, so $\T$ may be positive or negative; 
however setting
$\T\equiv\T_L=|\kpara|\vth\sqrt{3/2}|\hat{g}_2|^2$
(more generally,
$\Gamma_m\equiv\Gamma^L_m=|\kpara|\vth\sqrt{(m+1)/2}|\g|^2$)
amounts to an effective Landau-fluid-style closure that captures free-energy dissipation by Landau damping.
While not exact \cite{Kanekar14}, it
yields spectra in excellent agreement with %those found in 
linear drift-kinetic simulations \cite{ParkerThesis}.

\begin{figure}
  \includegraphics[]{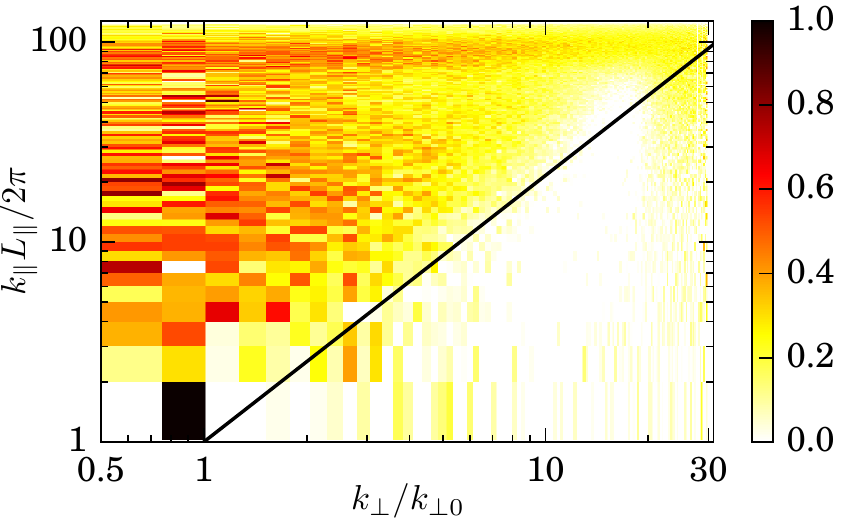}
  \caption{%}
	\label{fig:FluxSuppressionM2}
	$\ta{\bar{\Gamma}_2}=\ta{\T/\T_L}$, the free-energy transfer from fluid to kinetic moments in the saturated turbulent state, normalized to its value in a linear plasma,
 expressed in Fourier space.
		%Free-energy transfer to fine velocity-space scales is strongly suppressed.
		The line is $\taunli=\tausi$ (defined in the text).
  }
\end{figure}

We now consider the saturated state of the turbulence, in which the time averages $\ta{\tfd{\Wfluid}{t}}=\ta{\tfd{\Wkinetic}{t}}=0$,
so $\ta{\N}+\ta{\T}=\ta{\C}\geq0$. 
For linear phase mixing (and Landau damping) to play a similar role in a turbulent plasma as in a linear plasma, $\T$ would have to be similar to its linear value, $\ta{\T}\sim {\T_L}$.
In \fig~\ref{fig:FluxSuppressionM2}, we plot the ratio $\ta{\T/\T_L}$
% for saturated drift kinetic turbulence, Fourier-transformed to be 
as a function of $(\kperp,\kpara)$ %; % not $(\rperp,\kpara)$; 
(it is isotropic in $\rperp$).
The transfer is almost completely suppressed compared to the linear case, $\ta{\T/\T_L}\ll 1$, across a large range of wavenumbers that we will shortly characterize.

\newsection{Phase-mixing and anti-phase-mixing modes}%
The suppression of free-energy transfer associated with phase mixing is a nonlinear, kinetic effect.
To understand the suppression mechanism,  we decompose the distribution function into propagating modes in Hermite space,
by writing
$\g=(-i\ \sgn\ \kpara)^{m}\left[ \g^+ + (-1)^m \g^- \right]$,
where 
$\g^{\pm}=\tfrac{1}{2}(\pm i\ \sgn\ \kpara)^m\left[\g \pm i\ \sgn(\kpara) \gp\right]$.
The ``phase-mixing mode'', $\g^+$, propagates forward from low to high $m$,
the ``anti-phase-mixing mode'', $\g^-$, propagates backward from high to low $m$ \cite{Kanekar14,Parker15,Schekochihin15}.
For $m\geq3$, $|\g^{\pm}|^2$ evolves as
\begin{equation}
  \begin{split}
	&	\pd{}{t} \frac{|\g^{\pm}|^2}{2}
		\pm \frac{|\kpara|\vth}{\sqrt{2}} \pd{}{m}  \sqrt{m}|\g^{\pm}|^2 + \nu m^n|\g^{\pm}|^2
  \\
  &   
\hspace{0.5cm}
		= -\Re\Big\{\!\!\!\!\!\sum_{\ppara+\qpara=\kpara}\!\!\!\!\! [\g^{\pm}(\kpara)]^*
  \\
  & \hspace{1.1cm} \uperphat(\ppara)\ \cdot\nabla_{\perp}\left[ \delta^+_{\kpara\qpara}\g^{\pm}(\qpara) + \delta^-_{\kpara\qpara}\g^{\mp}(\qpara)\right]\Big\}, 
%%%    \label{eq:FreeEnergyEquationPlusMinus}
    \label{eq:GyrokineticEquationPlusMinus}
  \end{split}
\end{equation}
where $\delta^{\pm}_{\kpara\qpara}=\left[1\pm\sgn(\kpara\qpara)\right]/2$
($\delta^+_{\kpara\qpara}$ picks out $\kpara$ and $\qpara$ that have the same sign, $\delta^-_{\kpara\qpara}$ the opposite sign).
We have introduced a continuous approximation for the finite difference in $m$, valid because 
$\g^{\pm}$ are smooth in the sense that $\g^{\pm}\approx \hat{g}^{\pm}_{m+1}$ to lowest order at large $m$ \cite{Schekochihin15}.
In terms of $\g^{\pm}$, 
the normalized free-energy transfer from $m$ to $m+1$ is
\begin{equation}
  \begin{split}
		\bar{\Gamma}_m 
		\equiv \frac{\Gamma_m}{\Gamma^L_m} 
		&= \frac{\kpara\vth\sqrt{(m+1)/2}\ \Im(\gp^*\g)}{\kpara\vth\sqrt{(m+1)/2}\ |\g|^2}
		\\
		%&= \frac{|\g^+|^2-|\g^-|^2}{|\g^+|^2+|\g^-|^2}.
		&\approx \frac{|\g^+|^2-|\g^-|^2}{|\g^+|^2+|\g^-|^2},
    \label{eq:FreeEnergyTransferNormalized}
  \end{split}
\end{equation}
where $|\g|^2\approx|\g^+|^2+|\g^-|^2$ for large $m$ \cite{Schekochihin15}.
Any suppression of free-energy transfer, $\bar{\Gamma}_m<1$, can only be due to the anti-phase-mixing modes, $\g^-\neq0$.

Let us now consider the linear and nonlinear regimes in terms of $\g^{\pm}$.
In the linear case, we neglect the \rhs\ of \eqref{eq:GyrokineticEquationPlusMinus}.
Taking an initial disturbance in the fluid moments (low $m$) that propagates forwards only, we seek steady solutions with $\g^-=0$ (any $\g^-$ present in the initial conditions reflects off the hard-wall-like boundary condition at $m=0$ and becomes forward propagating within one streaming time).
The steady solutions of \eqref{eq:GyrokineticEquationPlusMinus} are
\begin{equation}
  \begin{split}
		%|\g^+|^2=\frac{A(\k)}{\sqrt{m}}\exp\left( \left(\frac{m}{m_c}\right)^{3/2}\right), 
%%%		|\g^+|^2=\frac{A(\kpara)}{\sqrt{m}}e^{\left(-{m}/{m_c}\right)^{3/2}}, 
%%%		\hspace{0.7cm}
%%%		m_c = \left(\frac{3|\kpara|\vth}{2\sqrt{2}\nu}\right)^{2/3},
		%|\g^+|^2=\frac{A(\kpara)}{\sqrt{m}}e^{\left(-{m}/{m_c}\right)^{n+\frac{1}{2}}}, 
		|\g^+|^2 &=\frac{A(\kpara)}{\sqrt{m}}\exp\left[{\left(-{m}/{m_c}\right)^{n+\frac{1}{2}}}\right], 
		\\ %\hspace{0.7cm}
		m_c &= \left[\frac{(n+1/2)|\kpara|\vth}{\sqrt{2}\nu}\right]^{1/(n+1/2)},
    \label{eq:GPlusLinearSolution}
  \end{split}
\end{equation}
where $A(\kpara)$ is an arbitrary function of $\kpara$ and $m_c$ is the collisional cutoff \cite{Zocco11,Kanekar14}.
For $m\ll m_c$, $|\g^+|^2$ has a $m^{-1/2}$ spectrum, while for $m\gtrsim m_c$, it is strongly damped.
As $\g^-=0$, the normalized Hermite flux \eqref{eq:FreeEnergyTransferNormalized} is $\bar{\Gamma}_m=1$. 

\begin{figure}
  \includegraphics[]{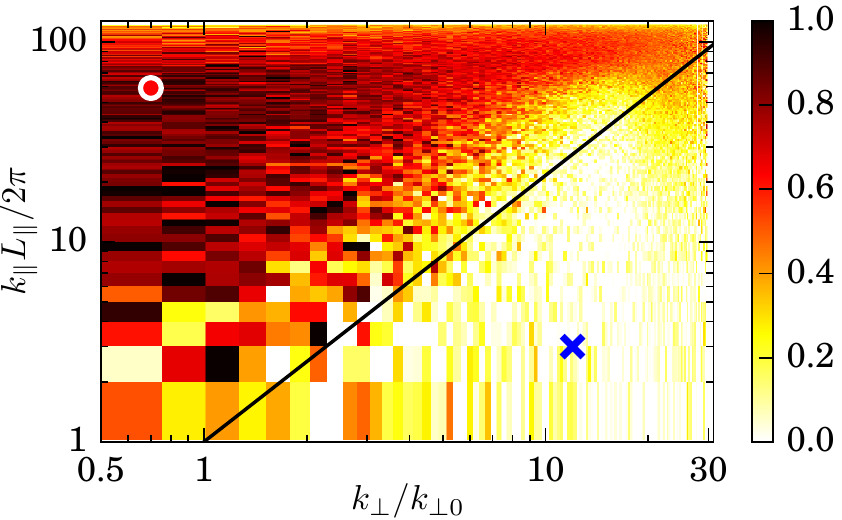}
  \caption{%}
    \label{fig:FluxSuppressionM100} 
    $\ta{\bar{\Gamma}_{30}}=\ta{\Gamma_{30}/\Gamma^L_{30}}$, free-energy transfer from mode $30$ to $31$, normalized to its value in a linear plasma, expressed in Fourier space. 
The line is $\taunli=\tausi$. 
%Free-energy transfer and hence Landau damping is completely suppressed at the energetically-dominant scales, for which $\taunli\gtrsim\tausi$. 
The marks show the locations of the Hermite spectra in \fig~\ref{fig:HermiteSpectra}.
  }
\end{figure}

To return to nonlinear drift kinetics, we reinstate the nonlinear term in \eqref{eq:GyrokineticEquationPlusMinus}.
Now, even if $\hat{g}^-_m$ is zero initially, the nonlinear term acts as a source in the ``$-$'' equation.
This puts some free energy into the $\g^-$ modes, with the result that $\bar{\Gamma}_m<1$, as in \fig~\ref{fig:FluxSuppressionM2}.
The effect is even clearer 
at large $m$, away from driving and dissipation: see \fig~\ref{fig:FluxSuppressionM100}, where we plot $\bar{\Gamma}_{30}$.
There are clear regions in wavenumber space where free-energy transfer is as in the linear case, $\bar{\Gamma}_{30}\approx1$,
and regions where free-energy transfer is completely suppressed, $\bar{\Gamma}_{30}\ll1$.

\newsection{Critical balance}%
To understand this partition of wavenumber space, we compare the characteristic rates of the phase mixing and nonlinearity,
respectively through the streaming rate $\tausi\sim \kpara\vth$ and the eddy turnover rate $\taunli\sim\kperp u_{\perp}\sim(\vth/L_{\parallel})(\kperp/\kperpzero)^{4/3}$ \cite{BarnesEtal11}.
When $\tausi\gg\taunli$, streaming dominates nonlinearity and the problem is essentially linear with $\bar{\Gamma}_m=1$. 
When $\tausi\ll\taunli$, nonlinearity dominates streaming and 
$\bar{\Gamma}_m<1$.
As shown in \figs~\ref{fig:FluxSuppressionM2} and \ref{fig:FluxSuppressionM100}, the line of critical balance, $\taunli=\tausi$, is in good agreement with the boundary of complete suppression $\bar{\Gamma}_m=0$, indicating that free-energy transfer is suppressed whenever $\taunli\gtrsim\tausi$.

\begin{figure}
  \includegraphics[]{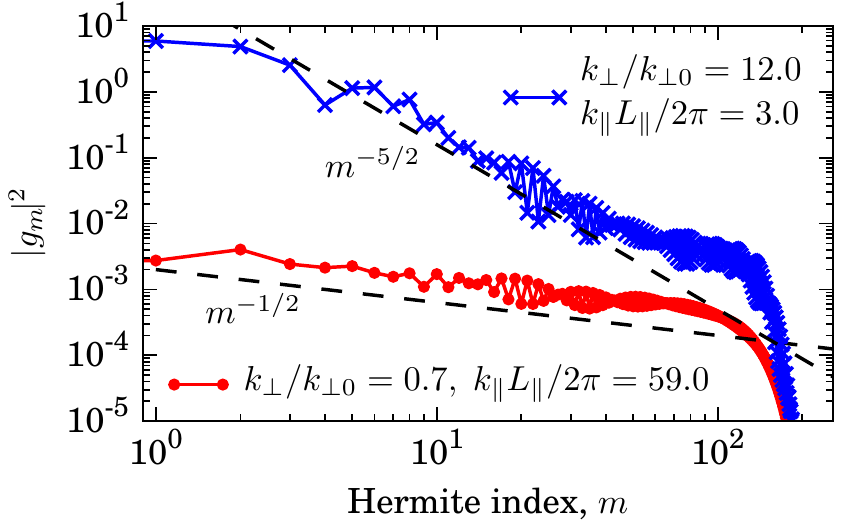}
  \caption{%}
    \label{fig:HermiteSpectra} Hermite spectra in the linear-streaming- ($m^{-1/2}$) and nonlinearity- ($m^{-5/2}$) dominated regions, for fixed $\kpara$ and $\kperp$.
  }
\end{figure}

The region $\taunli\gtrsim\tausi$ corresponds to the scales that contain the majority of the free energy (see \cite{ParkerThesis,Schekochihin15} for the theoretically predicted and numerically measured spectra). 
This has two important consequences.
Firstly, each velocity scale $m$ is, statistically, very nearly energetically decoupled from other scales, so, in particular, the fluid and kinetic moments are decoupled. 
Secondly, since $\sum_{\kpara}\int\d^2\rperp\ta{\N}=0$, 
the time average of \eqref{eq:FreeEnergyConservation} implies that the collisional dissipation rate, $\sum_{\kpara}\int\d^2\rperp\ta{\C}=\sum_{\kpara}\int\d^2\rperp\ta{\T}$, is also strongly suppressed, 
so collisional dissipation at fine velocity-space scales is a far less effective dissipation channel than in the case of linear Landau damping.
We confirm this suppression of collisional dissipation by considering the Hermite spectra.

\newsection{Hermite spectra and dissipation}%
The different free-energy transfer behaviors in the phase-mixing-dominated and the nonlinearity-dominated regions gives rise to two different Hermite spectra, plotted in \fig~\ref{fig:HermiteSpectra}
for fixed wavenumbers.
In the phase-mixing-dominated region ($\Gamma_m=1$), we observe the linear $m^{-1/2}$ spectrum \eqref{eq:GPlusLinearSolution}.
In the nonlinearity-dominated region ($\Gamma_m\ll1$), we observe the steep $m^{-5/2}$ spectrum predicted in \cite{Schekochihin15}.
The spectrum in the phase-mixing region gives rise to free-energy dissipation at the usual Landau damping rate:
for a fixed Fourier mode in this region,
		$\int_{3}^{m_c}\d m ~  \nu m^n m^{-1/2} 
		\sim |\kpara|\vth$.
This remains finite as $\nu\to0^+$.
In contrast, the dissipation rate for the $m^{-5/2}$ spectrum observed in the nonlinear region is
		$\int_{3}^{m_c}\d m ~  \nu m^n m^{-5/2} 
		\sim \nu^{4/3} (|\kpara|\vth)^{-1/3} \to0$
	as $\nu\to0^+$.
	Landau damping is thus suppressed in the nonlinear region.
As most of the free energy is contained in this region \cite{Schekochihin15,ParkerThesis},
	the total dissipation via phase mixing to collisional scales in $\vpara$ tends to zero as $\nu\to0^+$. 
	The vast majority of free energy cascades nonlinearly to dissipate at fine physical-space scales instead.
	
	These spectra and dissipation patterns explain the numerical observations of \cite{HatchEtAl13,Hatch14} that the Hermite spectrum summed over all Fourier space is much steeper than the linear $m^{-1/2}$ spectrum, and that free-energy dissipation via collisions in the inertial range decreases as $\nu$ decreases.
	Those observations may be understood as the result of aggregating the behaviors of
	the two distinct regions of Fourier space identified above.
	It is possible to prove that the aggregate Hermite spectrum is $m^{-2}$ \cite{Schekochihin15}, consistent with the scaling reported in \cite{HatchEtAl13,ParkerThesis}.

\newsection{Discussion}%
In this Letter, we have shown that linear phase mixing and the nonlinear cascade are strongly interdependent.
The nonlinear cascade in the inertial range excites anti-phase-mixing modes, suppressing the net transfer of free energy into kinetic modes.
This has both theoretical and practical implications. 
  Theoretically, these results profoundly change our understanding of the way in which free energy is cascaded and dissipated in phase space.
  As there is only a small net free-energy flux out of fluid modes in the inertial range, it is probably legitimate to neglect parallel streaming when deducing physical-space spectra from Kolmogorov arguments, as was done in \cite{BarnesEtal11}. 
  Since the Hermite spectrum at energetically-dominant scales is a steep $m^{-5/2}$ power law, these scales experience no free-energy dissipation via Landau damping as $\nu\to0^+$, and almost all free energy cascades to perpendicular spatial sub-Larmor scales.  
The steep Hermite spectrum also means that free-energy dissipation via linear phase mixing is not independent of collision frequency.
This has the important practical implication that enlarged collision frequencies cannot necessarily be used to compensate for low $\vpara$ resolution in weakly collisional simulations.

In this work, we have studied ITG drift-kinetic turbulence. 
However, our focus is on inertial-range physics, which does not depend on the details of the energy injection.
The approach presented here should be applicable to other kinetic systems where a nonlinearity interacts with particle streaming.
Indeed, similar suppressions of phase mixing has already been observed due to different nonlinearites in the Vlasov--Poisson system \cite{Parker15} 
and in kinetic passive scalar simulations \cite{KanekarThesis}.

\begin{acknowledgments}
The authors are grateful for fruitful conversations with 
I.~Abel, 
M.~Barnes,
G.~Colyer, 
S.~Cowley,
M.~Fox,
G.~Hammett,
F.~Parra,
C.~Roach, 
and
F.~van Wyk,
and especially 
A.~Kanekar
and
W.~Dorland.
This work was supported by the UK Engineering and
Physical Sciences Research Council through a Doctoral Training Grant award to J.T.P. and an Advanced Research
Fellowship [grant number EP/E054625/1] to P.J.D., with additional support from Award No KUK-C1-013-04 made
by King Abdullah University of Science and Technology (KAUST). Some of the results of this research were obtained using the
PRACE-3IP project (FP7 RI-312763), resource FIONN based in Ireland at 
the DJEI/DES/SFI/HEA Irish Centre for High-End Computing (ICHEC). This work also made use of the IRIDIS High Performance Computing Facility provided by the Science \& Engineering South (SES) Centre for Innovation, the UK HECToR HPC facility [grant
number EP/H002081/1], the resources of the STFC Hartree Centre,
the HELIOS supercomputer (IFERC-CSC), Admori, Japan,
and the University of Oxford Advanced Research Computing (ARC) facility \cite{Richards15}.
E.G.H's work has been carried out within the framework of the EUROfusion Consortium and was supported by a EUROfusion fusion researcher fellowship [WP14-FRF-CCFE/Highcock]. The views and opinions expressed herein do not necessarily reflect those of the European Commission. 
All authors are grateful to the Wolfgang Pauli Institute, Vienna, for its hospitality on several occasions.
\end{acknowledgments}

%\bibliography{itg_new}

%

\end{document}